\documentclass[twocolumn,secnumarabic,amssymb, nobibnotes, aps, prd]{revtex4-1}

\usepackage{graphicx}

\begin{document}

\title[]{Direct observation of the magnetic proximity effect in amorphous exchange-spring magnets by neutron reflectometry}

\author{A. J. Qviller}       %
\thanks{Corresponding author}%
\email{atlejq@gmail.com}        %
\affiliation{Institute for Energy Technology, P.O. Box 40, NO-2027 Kjeller, Norway}

\author{F. Magnus}%
\affiliation{Division for Materials Physics, Department of Physics and Astronomy, Uppsala University, Box 516, SE-751 20 Uppsala, Sweden}
\affiliation{Science Institute, University of Iceland, Dunhaga 3, 107 Reykjavik, Iceland}

\author{B. J. Kirby}
\affiliation{Center for Neutron Research, NIST, Gaithersburg, Maryland 20899, USA}

\author{B. Hj\"{o}rvarsson}%
\affiliation{Division for Materials Physics, Department of Physics and Astronomy, Uppsala University, Box 516, SE-751 20 Uppsala, Sweden}

\author{C. Frommen}%
\affiliation{Institute for Energy Technology, P.O. Box 40, NO-2027 Kjeller, Norway}

\author{B. C. Hauback}%
\affiliation{Institute for Energy Technology, P.O. Box 40, NO-2027 Kjeller, Norway}

\date{\today}

\begin{abstract}

In this letter we report a direct observation of a magnetic proximity effect in an amorphous thin film exchange-spring magnet by the use of neutron reflectometry. The exchange-spring magnet is a trilayer consisting of two ferromagnetic layers with high $T_c$'s separated by a ferromagnetic layer, which is engineered to have a significantly lower $T_c$ than the embedding layers. This enables us to measure magnetization depth profiles at which the low $T_c$ material is in a ferromagnetic or paramagnetic state, while the embedding layers are ferromagnetic. A clear proximity effect is observed 7 K above the intrinsic $T_c$ of the embedded layer, with a range extending 50 \AA.

\end{abstract}

\maketitle

Prominent magnetic proximity effects can be obtained at magnetic interfaces \cite{White:1985fg}.  These are present in a variety of magnetic thin films and heterostructures, often having profound effects on the observed magnetic properties \cite{Manna:2014bp,Magnus:2016hr}. As an example, proximity effects in non-magnetic spacer layers separating two ferromagnets can give rise to long-range interlayer exchange coupling\cite{Gokemeijer:1997wg}, changes in ordering temperature\cite{Bovensiepen:1998fd} and/or non-oscillatory alignment of magnetic layers\cite{Lim:2013fe,Gottwald:2013jl}. Since layered magnetic structures are ubiquitous in modern technology, the understanding of magnetic proximity effects is of fundamental importance.

\begin{figure} [tb]
  \centering
  \includegraphics[scale=0.46]{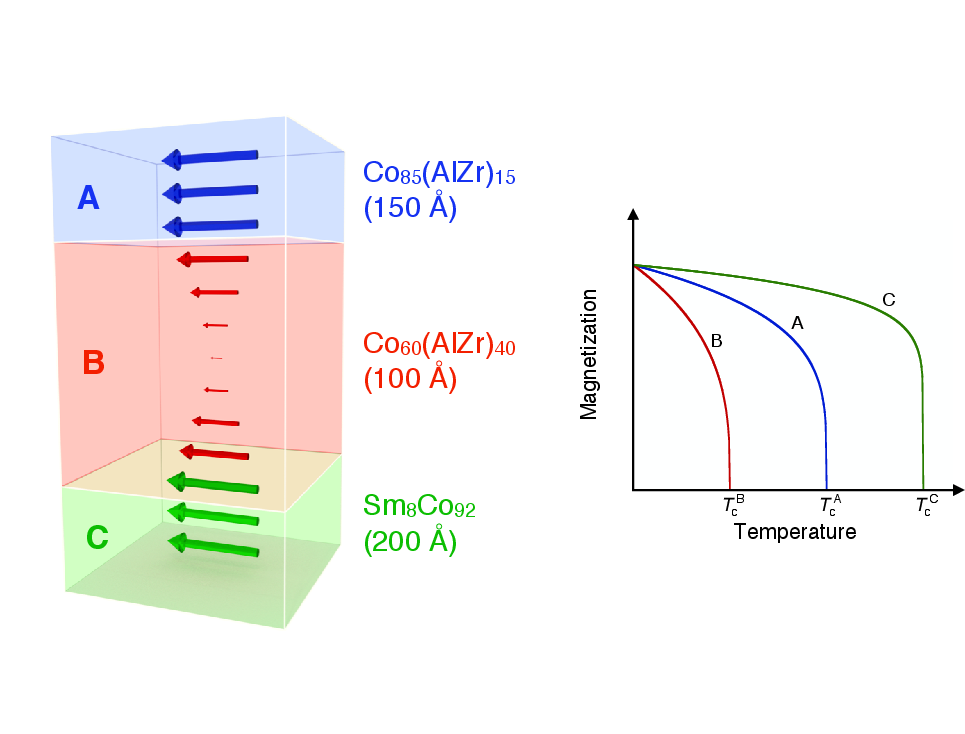}\\
  \caption{A schematic representation of the amorphous trilayer and an inferred magnetic proximity effect. The arrows denote the size of the magnetic moments. Layers A and C have a high magnetic ordering temperature whereas layer B has a low ordering temperature as shown on the right.  As illustrated in the figure, a pronounced proximity induced magnetization in layer B is expected at temperatures above $T_c^\mathrm{B}$. Figure adapted from reference.\cite{Magnus:2016hr} }
  \label{schematic}
\end{figure}

In this work we investigate the proximity effect in a ferromagnet-paramagnet system, more specifically in a trilayer \cite{Fullerton:1998gi} of an amorphous exchange-spring magnet\cite{Magnus:2016hr}. Amorphous heterostructures are free of step edges and grain boundaries and can therefore consist of well defined and smooth layers\cite{ChulMinChoi:2005br}. The exchange-spring magnet investigated here consists of three ferromagnetic layers, shown schematically in Fig.~\ref{schematic}. The top Co$_{85}$(AlZr)$_{15}$ layer (A) has an intrinsic $T_c$ well above room temperature and a small imprinted uniaxial anisotropy obtained as described in Ref.  \cite{Raanaei:2007, Magnus:2016hr}. The middle layer (B), which is magnetically isotropic Co$_{60}$(AlZr)$_{40}$, is engineered to have a $T_c$ well below room temperature, and much lower than the two other layers. The bottom layer (C) consists of Sm$_{8}$Co$_{92}$, which has a $T_c$ well above room temperature. Layer C has a large imprinted anisotropy, which can $e.g.$ increase the measured coercivity of the adjacent layers. This sample structure has previously been used to indirectly demonstrate that a proximity induced magnetization exists in layer B well above its intrinsic ordering temperature $T_c^\mathrm{B}$ \cite{Magnus:2016hr}. The inferred proximity effect was argued to result in exchange-spring behaviour at temperatures 50\% above $T_c^\mathrm{B}$ and exchange bias at even higher temperatures. However, no direct information concerning the magnetic state of the center layer was provided. Here we provide direct evidence of an induced magnetization in the low-$T_c$ middle layer and we also infer the magnetic profile throughout the layers, using polarized neutron reflectivity measurements \cite{Zabel:1994fo}. While proximity effects recently have been observed in multilayers \cite{Giant:2019} with the use of polarized neutron reflectivity, the trilayer sample in this work provides a cleaner and more exact measurement.




The samples were grown by dc magnetron sputtering in a UHV sputtering chamber at an Ar ($99.9999$ \% purity) sputtering gas pressure of $0.27$ kPa. First, a $20$ \AA$ $ thick seeding layer of Al$_{70}$Zr$_{30}$ was deposited on a Si(100) substrate (with the native oxide) from an Al$_{70}$Zr$_{30}$ alloy target of purity $99.9$\%. 
Subsequently, a $200$ \AA$ $ thick Sm$_{8}$Co$_{92}$ alloy film was grown by co-sputtering from€ elemental targets of Co ($99.9$\% purity) and Sm (99.9 \% purity), after which a Co$_{60}$(Al$_{70}$Zr$_{30}$)$_{40}$ of $100$ \AA$ $ and a Co$_{85}$(Al$_{70}$Zr$_{30}$)$_{15}$ layer of 150 \AA$ $ were grown by co-sputtering from the Co and AlZr targets. Finally, a $30$ \AA$ $ thick capping layer of Al$_{70}$Zr$_{30}$ was grown to protect the underlying magnetic trilayer from oxidation. All films were grown at room temperature. 
Two permanent magnets provided a magnetic field of approximately $0.1$ T parallel to the plane of the films during growth as described in Ref. \cite{Raanaei:2007}. This induces a uniaxial in-plane anisotropy in the ferromagnetic layers having an ordering temperature above the growth temperature (A and C). 
The atomic flux as a function of magnetron power was determined for each target material through X-ray reflectivity measurements of films grown from each of the magnetrons. The power on each magnetron was then set to achieve a given composition while co-sputtering. Rutherford backscattering measurements have previously confirmed that this is a robust method for the materials in question \cite{RBS1:2007, RBS2:2016}. MOKE measurements were carried out on the samples to confirm the $T_{c}$ of the middle layer and that the magnetization loops of the trilayers were consistent with previously studied samples. More details on the growth and structural characterization can be found in Refs.~\onlinecite{Magnus:2016hr}, ~\onlinecite{Magnus:2013ey} and ~\onlinecite{Raanaei:2007}.


Polarized neutron reflectivity experiments were carried out at the PBR beamline at NIST at a wavelength of $\lambda = 4.75$ \AA $ $ using an instrument resolution of $\Delta \lambda / \lambda = 0.01$. Four reflectivities, corresponding to the two non-spin-flip channels ($R^{++}$ and $R^{--}$) as well as the two spin-flip channels ($R^{-+}$ and $R^{+-}$), were measured out to $q = 0.2$ \AA $^{-1}$  at $T_{1} = 300$ K, $T_{2} = 110$ K and $T_{3} = 10$ K. With $T_c^\mathrm{B} = 103 \pm 1$ K \cite{Magnus:2016hr} (see Figure 1), these temperatures correspond to $T_{1} \gg T_c^\mathrm{B}$, $T_{2} > T_c^\mathrm{B}$ and $T_{3} < T_c^\mathrm{B}$. 
Samples were measured with an applied external field of $\mu_{0}H = 10$~mT along the easy axis of the imprinted anisotropy and the scattering plane perpendicular to this axis. Measurements of the spin-flip reflectivities returned mainly noise, consistent with the presence of a collinear magnetization state \cite{Zabel:1994fo}, which is reasonable in the given measurement configuration. The spin-flip reflectivities were therefore subsequently disregarded in the fitting process.

\begin{figure} [h!]
  \centering
  \includegraphics[scale=0.333]{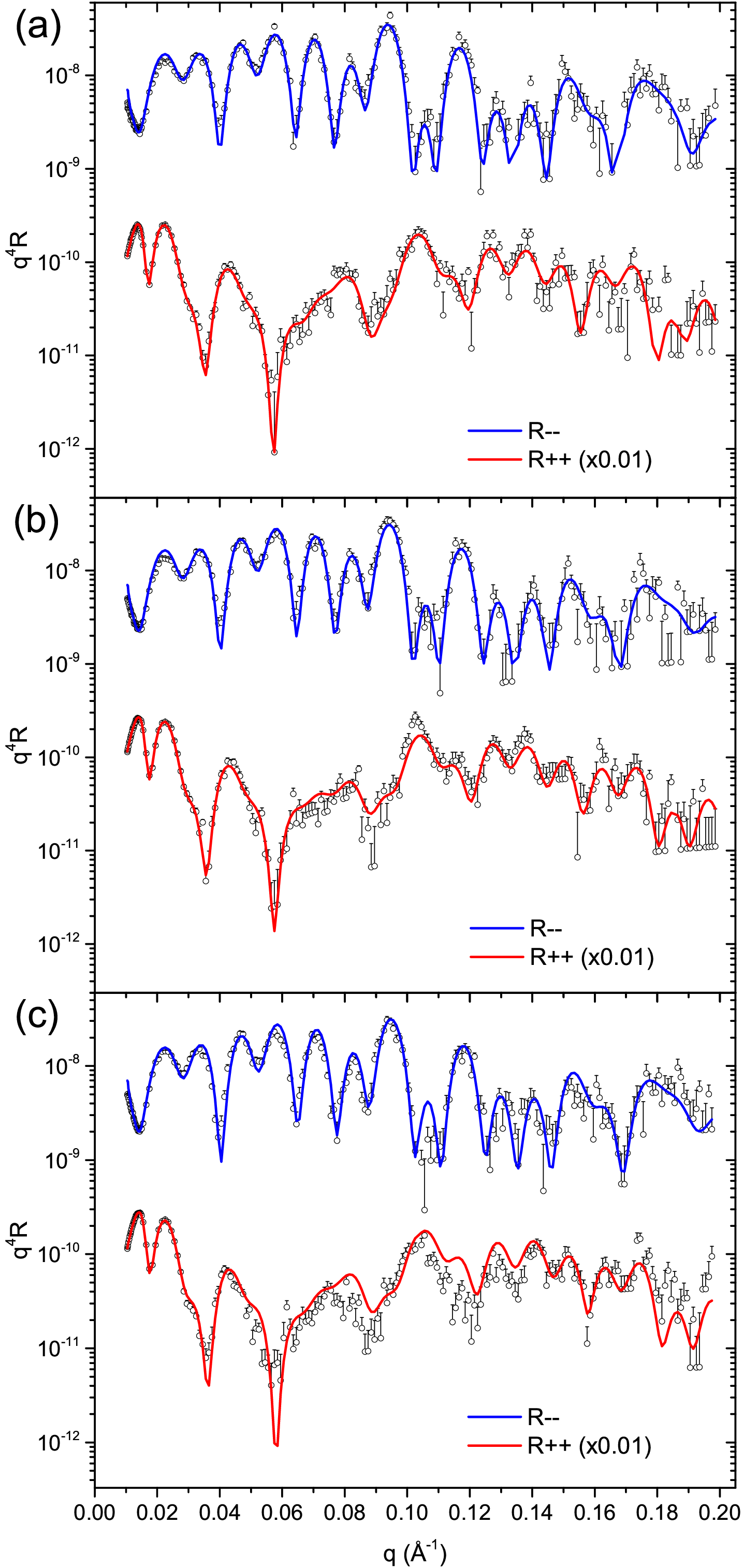}
  \caption{Non-spin-flip polarized neutron reflectivity scans, $R^{--}$ and $R^{++}$, measured at $\mu_{0}H = 10$ mT and (a) $T = 300$ K, (b) $T = 110$ K and (c) $T = 10$ K. Fits are shown as solid lines in red and blue. Error bars correspond to $\pm{1}$ standard deviation and are shown only on the upper side for clarity.}
  \label{f2}
\end{figure}



Data were fitted using the GenX 2.4.7 reflectivity package with the new MagRefl module \cite{Bjorck:2007eb}, using the logarithm of the reflectivity as the figure of merit. The sample model consists of a partially oxidised capping layer (oxide consisting of a 70/30 mixture of Al$_{2}$O$_{3}$/ZrO$_{2}$ and and a Al$_{70}$Zr$_{30}$ layer), a magnetic Co$_{85}$(AlZr)$_{15}$ layer, a magnetic Co$_{60}$(AlZr)$_{40}$ interlayer, a magnetic Sm$_{8}$Co$_{92}$ layer and a seeding layer of Al$_{70}$Zr$_{30}$ on a thin SiO$_{2}$ layer on a Si substrate (see Figure 3). Most structural parameters were determined by fitting the results obtained at $T = 300$ K and $\mu_{0}H = 10$ mT. Reflectivity measurements cannot be used alone to determine the chemical composition of layers consisting of more than two elements, but the validation of the sample preparation procedure stated above justifies fixing the stoichiometry of the layers to the intended values and only allowing their densities to vary during the fitting at $T = 300$ K and $\mu_{0}H = 10$ mT. In the simulations, the low-$T_c$ layer was defined by $10$ slices and the magnetic moments of these layers were fitted to a sum of two power laws with the same exponent, one corresponding to a decaying magnetization from each interface, induced by the neighbouring ferromagnetic layers as illustrated in Fig. \ref{schematic}. A power law decay of the magnetization was chosen as it is the functional form of the long-range exchange interaction, as $\it{e.g.}$ described in Refs. ~\onlinecite{Magnus:2016hr} and \onlinecite{Fisher:1972}. The resulting step-wise magnetic profile was smoothed by allowing a small, linked chemical roughness (7 \AA $ $ RMS) for each slice. 
Only the magnetization of the three ferromagnetic layers and their profiles were allowed to vary when fitting the data obtained at $T = 110$ K and $T = 10$ K.
Additionally, changes in thickness were needed to account for the thermal expansion in the layers, which was determined to be
4.1 x 10$^{-5}$ K$^{-1}$, or $1.2$ \% when heating the sample from $T = 10$ K to $T = 300$ K. Notice that the determined thermal expansion only holds for the combined film and substrate, where the substrate provides elastic boundaries defining the changes in the lateral direction with temperature, due to clamping effects. The resulting polarized neutron reflectivity results (scaled in $q$) and corresponding fits are shown in Fig. \ref{f2} (a), (b) and (c).

\begin{figure} [h]
  \centering
  \includegraphics[scale=0.333]{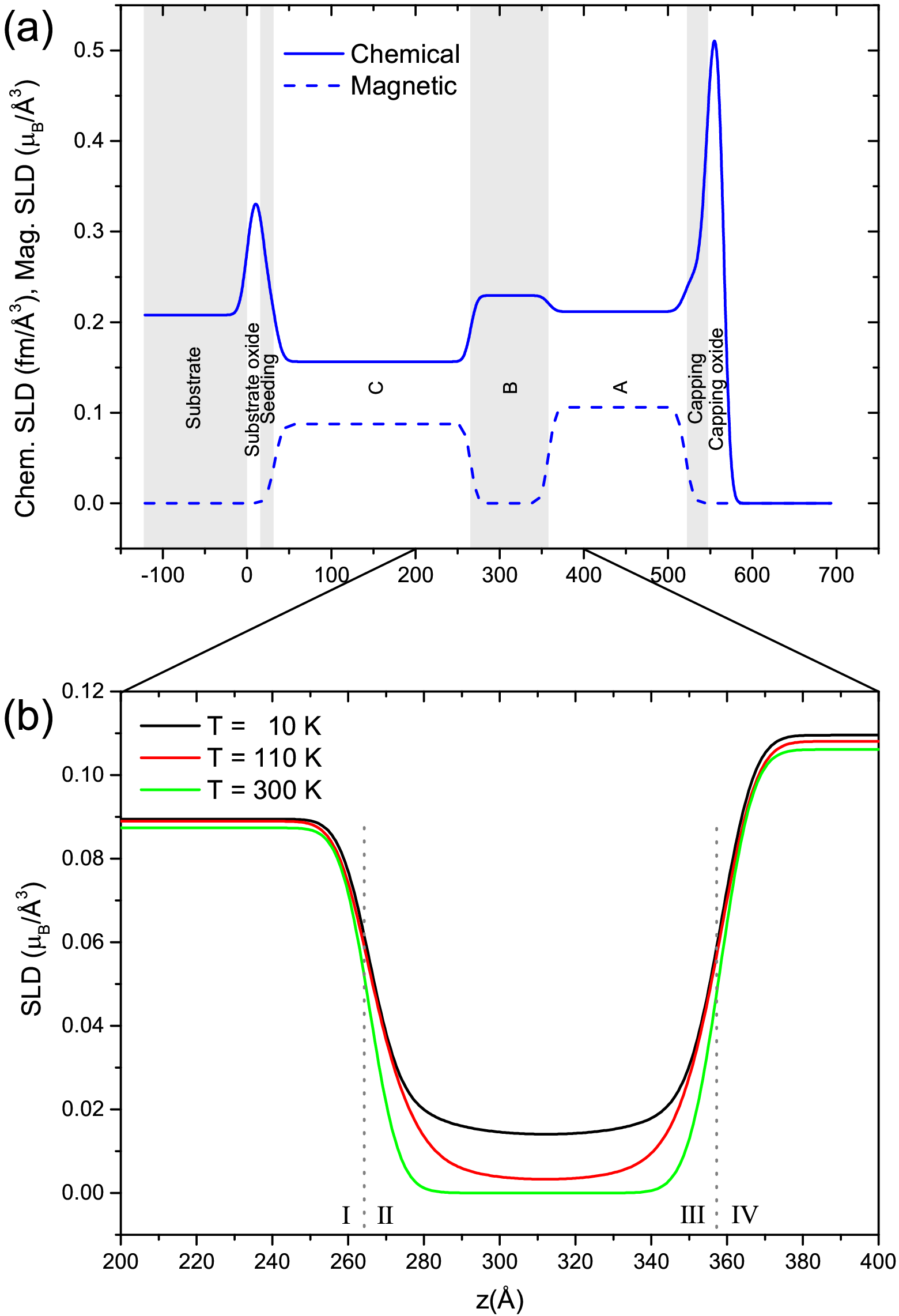}\\
  \caption{(a) Chemical and magnetic SLD profiles at $\mu_{0}H = 10$ mT and $T = 300$ K for the entire exchange-spring magnet heterostructure. The capping layer, layers A, B, C, the seeding layer and the substrate are indicated. Oxide layers on the capping layer and substrate are also marked. (b) Magnetization profiles of the low-$T_c$ Co$_{60}$(AlZr)$_{40}$ middle layer and its interfaces at $\mu_{0}H = 10$ mT, $T = 10$, $110$ and $300$ K. Error bars (not shown) of the magnetization in the center of this layer are 7\% for all temperatures. The magnetization profiles at $T = 10$ K and $110$ K have been offset to correct for the measured thermal expansion.}
  \label{f3}
\end{figure}

The determined chemical and magnetic SLD profiles at $\mu_{0}H = 10$ mT and $T = 300$ K are shown in Fig. \ref{f3} (a), as functions of the distance from the substrate $z$.
The chemical SLD profile yields interface roughnesses equal to or less than $8$ \AA$ $ RMS. Inferred magnetic SLD profiles are shown in Fig.~\ref{f3} (b). When $T = 300$ K and $\mu_{0}H = 10$ mT (green line), no magnetization is observed in the middle of layer B. Furthermore, the magnetization at the interfaces decays sharply, consistent with relatively short ranged magnetic proximity effects. At the lowest temperature, $T = 10$ K (black line), the middle layer is magnetized, as expected since this is well below the ordering temperature of layer B. Significant magnetization is seen throughout the middle layer at $T = 110$ K (red line) which is well above the ordering temperature of the layer  (7~K above $T_c^\mathrm{B}$). This observation can be viewed as a manifestation of a long range magnetic proximity effect, where a magnetization is induced in layer B well above its intrinsic ordering temperature due to the proximity to the high-$T_c$ layers A and C. Error bars for the magnetization in the middle of layer B as measured by reflectometry were determined by inspection of how much it can change before the fits become visually worse. This was 7\% for both the $T = 10 $ K data and the $T = 110 $ K data. For technical reasons, the latter error bar in absolute terms was also used for the interlayer magnetization at $T = 300 $ K.

As there is a significant magnetization in the center of the $100$ \AA$ $ thick layer B, it can be deduced that the range of the proximity effect is at least $50$ \AA$ $ at $T = 110$ K. This is consistent with the interpretation of previous experimental results and requires an explanation in terms of  long-range effective exchange interactions \cite{Magnus:2016hr}. Monte Carlo simulations have shown that an extended region of induced magnetization can only be expected in layer B, if interactions beyond nearest neighbour are accounted for. Furthermore, atomic correlations in terms of regions of higher Co-density can further amplify the proximity effect, as discussed in Ref. ~\onlinecite{Magnus:2016hr}. Interdiffusion can be ruled out as the cause of the induced magnetization in layer B due to the very different length scales of its range and the magnitude of the interface roughnesses, and that the measured interdiffusion can not depend on the measurement temperature.

\begin{table}[]
\centering
\caption{Absolute value of the half-width half maxima of the derivative of the magnetic SLD profiles.}
\begin{tabular}{|l|l|l|l|l|} \hline
 Temperature (K) &  I (\AA) & II (\AA) & III (\AA) & IV (\AA) \\ \hline
           10$\pm{1}$ & 7.3 & 8.7  & 9.1 & 8.3 \\ \hline
          110$\pm{1}$ & 8.3 & 12   & 12  & 8.6 \\ \hline
          300$\pm{1}$ & 7.4 & 7.3  & 9.1 & 7.9 \\ \hline
\end{tabular}
\label{t1}
\end{table}

To investigate the spatial dependence of the magnetization with temperature, the absolute value of the derivative of the magnetic SLD profiles in Fig. \ref{f3} (b) was calculated. Four half-widths at half maxima were extracted, describing the spatial change in magnetization at the interfaces of the layers. The results are provided in Table \ref{t1}, the annotations of the interfaces I-IV are defined in Fig. \ref{f3} (b). The widths I and IV describe the change in the magnetization of the outer boundaries of layers A and C. These are found to be independent of temperature, as expected.
The changes in width II (III) describes the changes in the magnetization profile of the interface of A (C) and B. At $T = 10$ K, all the layers are close to fully magnetic and the deduced magnetization profile resembles therefore primarily the distribution of the elements in the sample. 
At $T = 300$ K, the magnetization and susceptibility of layer B is negligible and the width is therefore dominated by the interface effects of layers A and C, at both the interfaces. Thus the width of the magnetic profile obtained at these temperatures are expected to be similar, which is consistent with the obtained results.
We can now compare these with the results obtained at $T = 110$ K. The width of the magnetic profile in region II and III is clearly larger at this temperature, as seen in Table \ref{t1}. In order to underpin the above analysis, a normal probability plot of the values in Table \ref{t1} was done (not shown). It revealed that the widths in region II and III at $T = 110$ K are outliers at 2.0 $\sigma$, while the rest of the observations are approximately normally distributed around a sample mean of 8.8 \AA. 
The above discussion reflects the expected changes in the magnetic susceptibility, which is largest at $T_{c}$. Monte Carlo simulations involving beyond nearest neighbour magnetic interactions show that not only is the peak in the magnetic susceptibility shifted to higher temperature, but it is also significantly broadened, due to proximity effects. \cite{Magnus:2016hr} This is well reflected in the present results, however, the magnetization of the centre of the layer is not captured by these simulations. An analogous effect arising from changes in the effective coupling due to changes in the local concentration has previously been described for random alloys \cite{Gemma:2018}. 
The spatially dependent changes in the concentration of the elements in a magnetic alloy will give rise to changes in the effective magnetic coupling. The magnetization will consequently decline in a non-uniform way, giving rise to partially coupled ferromagnetic regions. This, in combination with long range interaction can therefore serve as a basis of the understanding of the observed proximity effects, well above its intrinsic ordering temperature.

The Research Council of Norway is acknowledged for financial support through the SYNKN\O YT program, project 218418. This work was also funded by the Swedish Research Council (VR) and the Knut and Alice Wallenberg Foundation (KAW). FM acknowledges funding from the Icelandic Research Fund grant no. 174271-051. AJQ would like to express his gratitude to Bengt Lindgren for his skillful assistance with modelling in GenX.

%
%
%
%
%
%
%
%
%

\end{document}